\documentclass[aps,preprint,tightenlines,showpacs,nofootinbib]{revtex4}
\usepackage{epsfig}
\begin{document}
\title{Too many $X's$, $Y's$ and $Z's$?}
\pacs{14.40.-n,12.40.Yx}
\author{T.~F.~Caram\'es}
\affiliation{Departamento de F{\'\i}sica Fundamental,
Universidad de Salamanca, 37008 Salamanca, Spain}
\author{A.~Valcarce}
\affiliation{Departamento de F{\'\i}sica Fundamental,
Universidad de Salamanca, 37008 Salamanca, Spain}
\author{J.~Vijande}
\affiliation{Departamento de F\'{\i}sica At\'{o}mica, Molecular y Nuclear, Universidad de Valencia (UV)
and IFIC (UV-CSIC), Valencia, Spain.}
\date{\emph{Version of }\today}

\begin{abstract}
A large number of new states have been reported during the last few years in charmonium spectroscopy above 
the charmed meson production threshold. They have been called $X's$, $Y's$, and $Z's$.
We reflect on the influence of thresholds on heavy meson spectroscopy comparing different
flavor sectors and quantum numbers. The validity of a quark-model picture above
open-flavor thresholds would severely restrict the number of channels that may lodge meson-meson
molecules.  

\end{abstract}

\maketitle

There is general agreement among the hadron physics community that the constituent
quark model is a good phenomenology for charmonium and bottomonium below their
respective flavor thresholds. It is in this context that discoveries during the recent
years of new states in the heavy flavor sector provide tantalizing hints of the
underlying dynamics elsewhere in spectroscopy. If this was established it could lead
to a more unified and mature picture of hadron spectroscopy. A recurring old question
arises again: being some of these new states firmly established (as it is the case of 
the $X(3872)$), could they be fitted into the quark
model scheme or are we in front of the breakdown of such a pattern?

The question above has been the flagship of many experimental and theoretical
efforts during the last two decades. In spite of the success of the quark-model
picture for the description of the heavy meson spectra~\cite{Isg83}, the limit
for its applicability was questioned, trying to learn 
why the constituent quark model was wrong~\cite{Bal93}. The possible 
impact of thresholds on the hadronic spectrum came up soon as a relevant mechanism 
in an attempt to go beyond the adiabatic approximation~\cite{Isg99}.
The scarce experimental data did not allow to arrive to any definitive
conclusion. Such scenario was relived with the hectic times of the $\theta^+$.
The existence of this state was, for some time, considered a possible hint pointing
to the end of the constituent quark model~\cite{Clo03}. At the same time,
the first warnings about the experimental data were highlighted.
It was noticed that listing all of the mesons
from the PDG as a function of $J^{PC}$ indicated that the light hadron
dynamics was clearly overpopulated, showing that not all data could
be correct. The increasing experimental data in the excited
heavy-meson spectra came to reinforce the suspect that the successful 
quark-model picture for charmonium may get significant distortion
from the $D\bar D$ threshold region. On the experimental side, the need of caution
about the proliferation of low-statistics experimental data has been posed. 
It has been noticed that experimental signals of $3\sigma$ will probably
disappear 80\% of times~\cite{Qig11}. Moreover, there is a well-known difficulty for 
disentangling resonances from cusps due to the opening of thresholds, what may provide
an alternative explanation to some of the recently reported new states~\cite{Bug11}.

Our purpose in this letter is to discuss how the new set of states reported could offer
insight to check the validity of the constituent quark model beyond flavor thresholds.
It would severely restrict the number of channels that may lodge meson-meson
molecules. Out of the many states recently reported in charmonium and bottomonium 
spectroscopy, we do not really know how many will survive future experimental 
screenings. We are still shocked by the $\theta^+$ resonance, seen by so many experiments 
that later on did not find anything. There is only one state that has been
firmly established by different collaborations and whose properties seem 
to be hardly accommodated in the quark-antiquark scheme, this is the $X(3872)$~\cite{Cho03}.
Regarding the zoo of other states that have been proposed, we have to stay tuned
but also be cautious~\cite{Clo03,Qig11}. Some members of this $XYZ$ jungle are awaiting confirmation,
seen only by one collaboration, like the intriguing charged 
state $Z(4430)$, seen by Belle but not by BaBar~\cite{Cho08}. Other members of this jungle 
cannot be excluded to fit into the simple quark-antiquark scheme, like the
$Z(3930)$ recently identified as the $\chi_{c2}(2P)$ charmonium state~\cite{Aub10}. 
Other experimental signals seen only by a particular experiment, in some of the 
expected decay modes, or with low statistics, could just be the reflection of the opening
of thresholds~\cite{Bug11}. 

\begin{figure*}
\vspace*{-0.5cm}
\includegraphics[width=17cm,height=20cm]{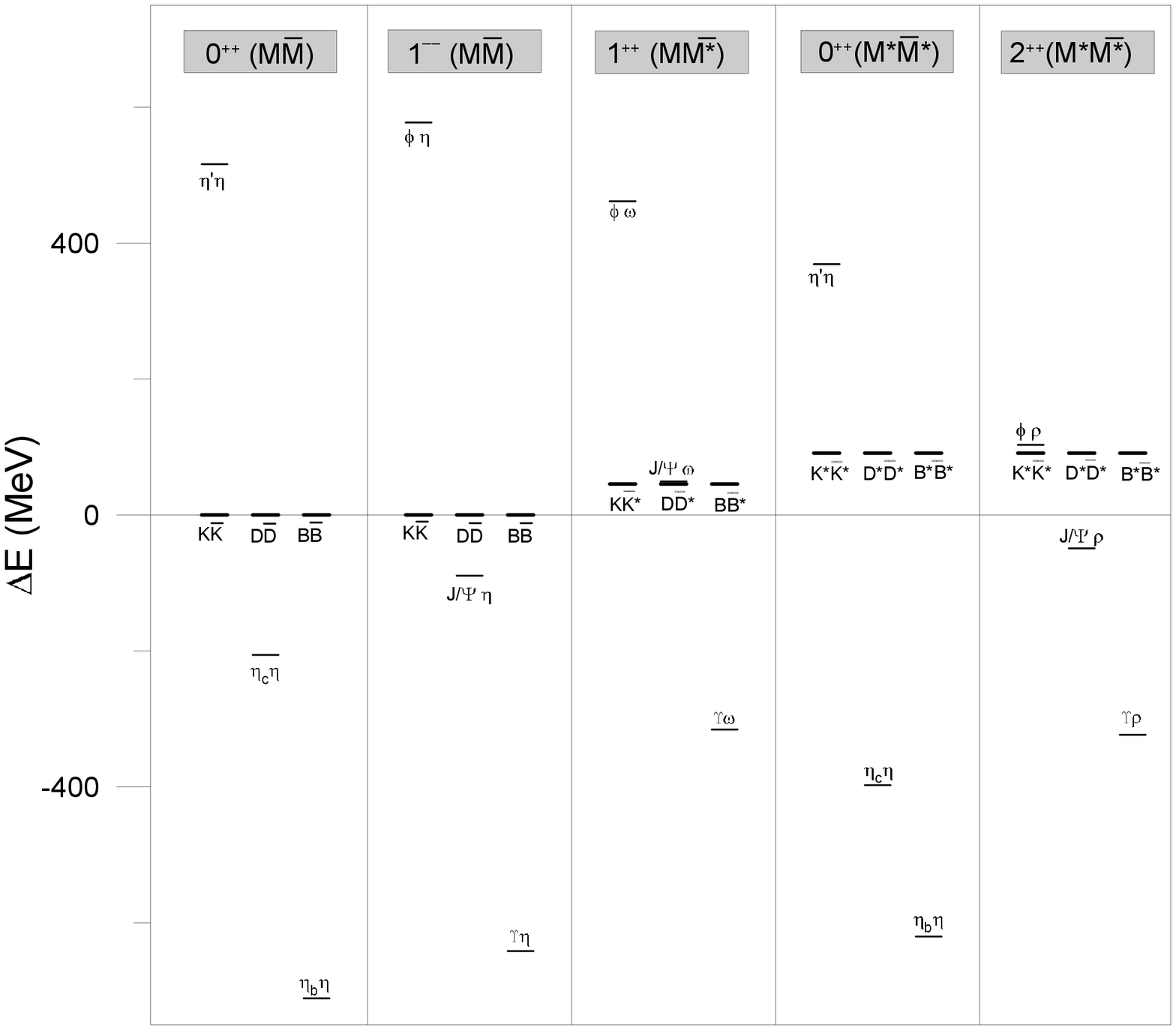}
\vspace*{-8.5cm}
\caption{Experimental thresholds of four-quark systems made of a heavy and a light
quark and their corresponding antiquarks, $Qn\bar Q \bar n$ with 
$Q=s$, $c$, or $b$, for several sets of quantum numbers, $J^{PC}$. We have set as 
our origin of energies the $K\bar K$, $D\bar D$ and $B\bar B$ masses for the hidden
strange, charm and bottom sectors, respectively.}
\label{fig1}
\end{figure*}

Let us start by discussing Fig.~\ref{fig1}. In this figure we have plotted 
the experimental thresholds~\cite{Pdg10} of four-quark systems made of a heavy and a light
quark and their corresponding antiquarks for several sets of quantum numbers, $J^{PC}$, in 
three different flavor sectors: $Q=s$, hidden strange; $Q=c$, hidden charm; and 
$Q=b$, hidden bottom. In every flavor sector we represent the mass difference 
with respect to the mass of $K\bar K$,
$D\bar D$ and $B\bar B$, respectively. In a constituent quark model picture, the four-quark state
$Qn \bar Q \bar n$ could either split into $(Q \bar n) - (n \bar Q)$ or
$(Q \bar Q) - (n \bar n)$. One observes how the general trend for all quantum 
numbers is that the mass of the $(Q \bar Q) - (n \bar n)$ two-meson system is 
larger than the mass of the $(Q \bar n) - (n \bar Q)$ two-meson state for $Q=s$, but it is 
smaller for $Q=c$ or $b$. It is remarkable the case
of $J^{PC}=1^{++}$ for $Q=c$, where the $(Q\bar Q) - (n\bar n)$ and 
the $(Q \bar n) - (n \bar Q)$ two-meson states are almost degenerate. The reverse of the ordering of
the masses of the $(Q \bar Q) - (n \bar n)$ and $(Q \bar n) - (n \bar Q)$
thresholds when increasing the mass of the heavy quark for all $J^{PC}$
quantum numbers can be simply understood within the constituent quark model
with a Cornell-like potential~\cite{Clo03}. 
The binding of a coulombic system is proportional to the reduced mass of the
interacting particles. Thus, for a two-meson threshold with a heavy-light light-heavy 
quark structure, the binding of any of the two mesons does not change much when increasing the mass 
of the heavy flavor, due to the reduced mass of each meson being close
to the mass of the light quark. However, if the two-meson state presents a
heavy-heavy light-light quark structure, the binding of the heavy-heavy meson increases with the mass of 
the heavy particle while that of the light-light meson remains constant,
becoming this threshold lighter than the heavy-light light-heavy two-meson structure,
as seen in Fig.~\ref{fig1}. 

\begin{figure*}[t]
\vspace*{-2cm}
\resizebox{7.cm}{10.cm}{\includegraphics{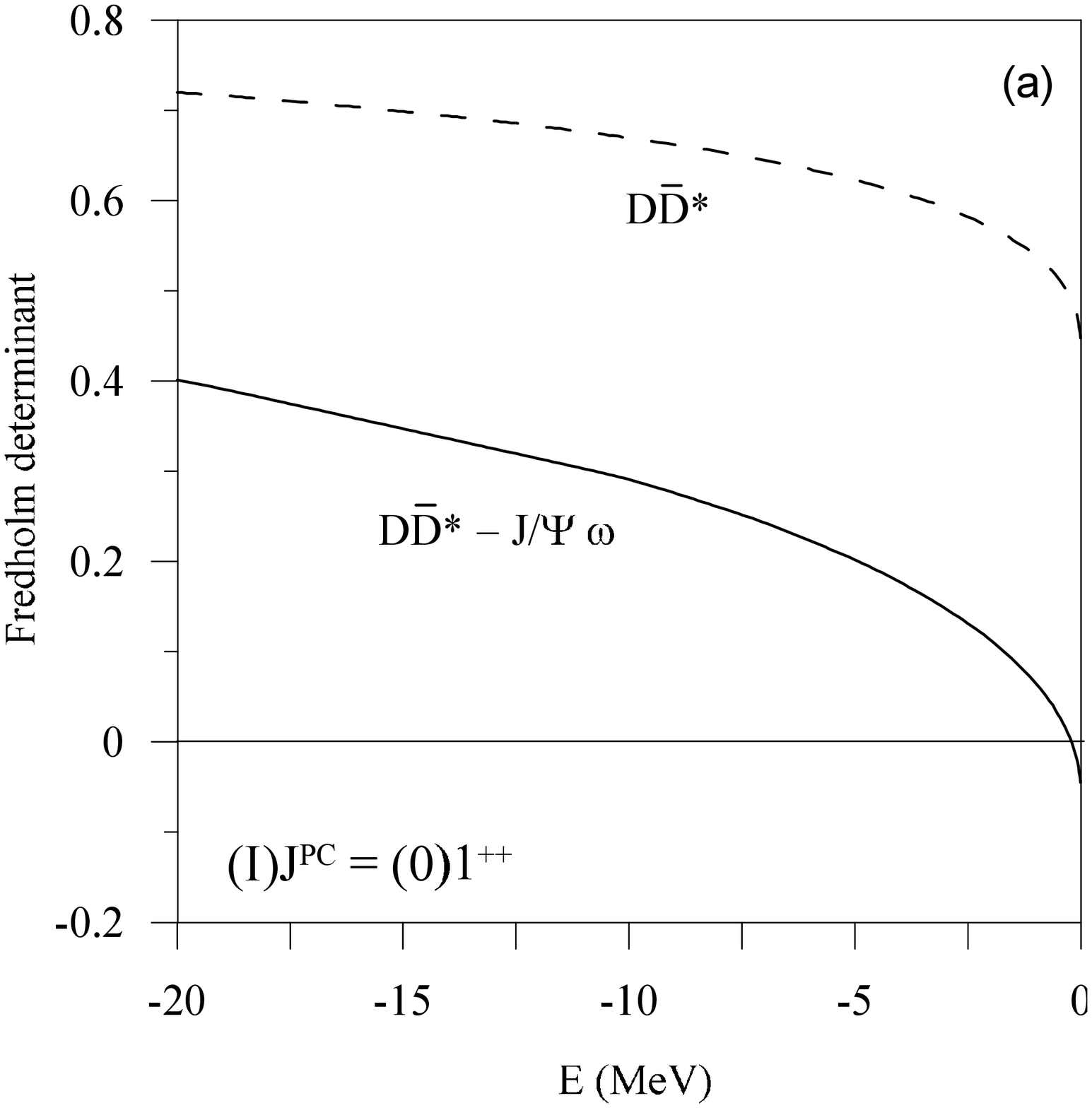}}\hspace{2cm}
\resizebox{7.cm}{10.cm}{\includegraphics{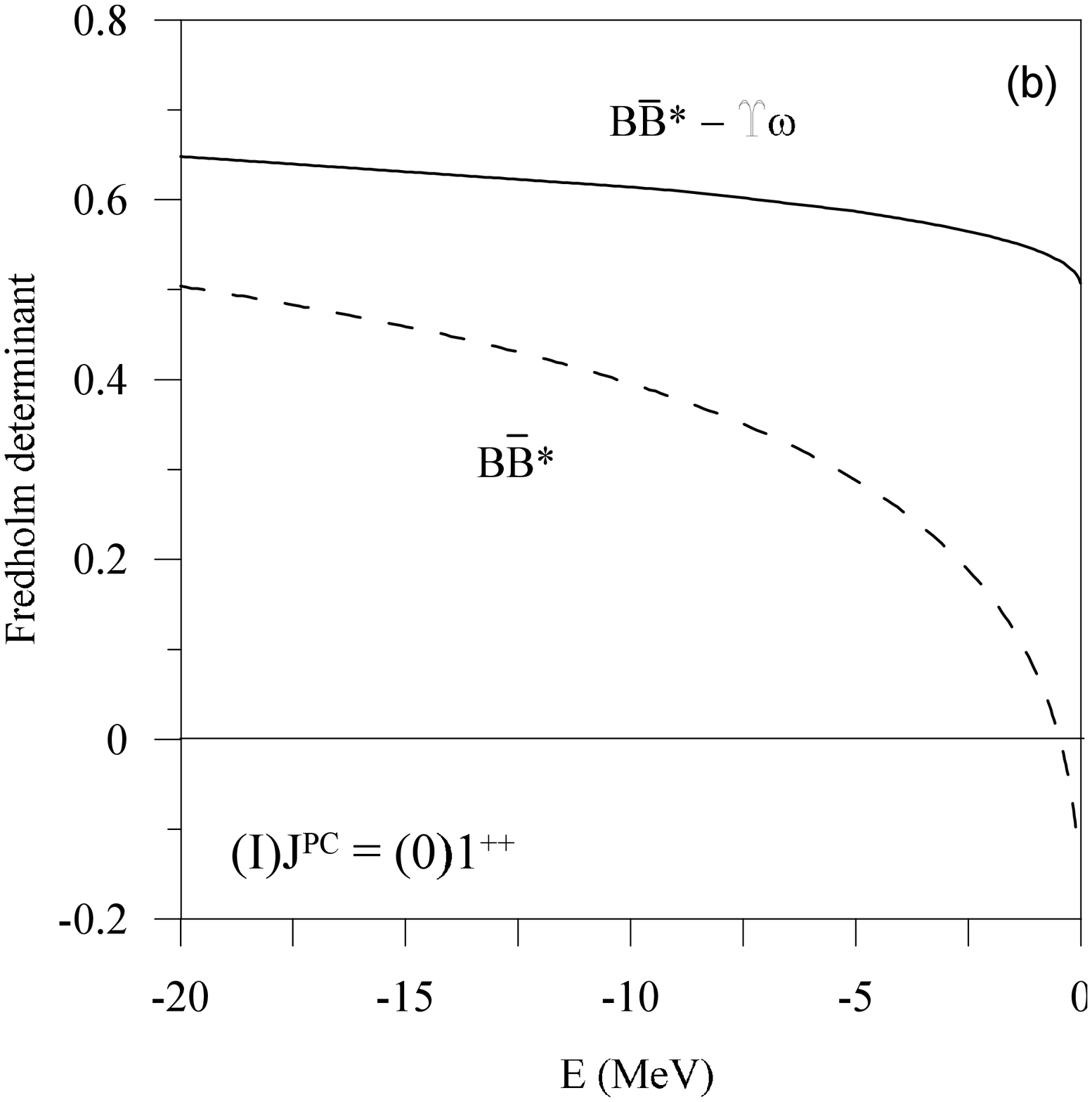}}
\vspace*{-2cm}
\caption{(a) $(I)J^{PC}=(0)1^{++}$ $cn\bar c\bar n$ Fredholm determinant~\protect{\cite{Car09}}. The dashed line 
stands for a calculation considering only charmed mesons, the solid line
includes also the $J/\Psi \omega$ two-meson system. (b) Same
as (a) for bottomonium.}
\label{fig2}
\end{figure*}

Such a picture, together with the absence of long-range
forces~\cite{Tor92} in a charmonium-light two-meson system,
may suggest different consequences for the existence of molecules
close to the meson-antimeson threshold.
First, the possible existence of such molecules in the 
hidden-strange sector. If the $K\bar K$ interaction is attractive for some particular set
of quantum numbers, this two-meson system may be stable because 
no any other threshold appears below, the dissociation of the molecule
being therefore forbidden (see the $Q=s$ states for
any $J^{PC}$ quantum numbers in Fig.~\ref{fig1}). Such a possibility would become
more probable for those quantum numbers where the quark model
seems to work worst, those cases where one needs a P-wave in the
simplest quark model structure, $q \bar q$, but can be obtained in
S-wave from a four-quark system. In these cases, the mass of the four-quark system
could be even below the predicted lowest quark-antiquark state.
This is precisely the idea suggested by Weinstein and
Isgur~\cite{Wei90} as a plausible explanation of the proliferation of 
scalar mesons in the light quark sector. They concluded the $J^{PC}=0^{++}$ and $1^{++}$
quantum numbers to be the best candidates to lodge a meson-antimeson molecule.
These quantum numbers are P-wave in the quark model but S-wave in the four-quark picture
and besides they are spin triplet, having therefore an attractive spin-spin 
interaction~\cite{Wei90}. 

Second, the possibility of finding meson-antimeson molecules contributing to 
the meson spectrum becomes more and more difficult when increasing the mass
of the heavy flavor, due to the lowering of the mass of the $(Q \bar Q) - (n\bar n)$ 
threshold (see the $Q=c$ or $b$ states for any $J^{PC}$ quantum numbers in Fig.~\ref{fig1}).
This would make the system dissociate immediately. In such cases,
the presence of attractive meson-antimeson thresholds would manifest in the scattering cross section but 
they will not lodge a physical resonance. These ideas favored 
the interpretation of several of the experimental
signals in charmonium and bottomonium spectroscopy above flavor thresholds as originated from the 
opening of the threshold and not being resonances~\cite{Bug11}. 

\begin{figure*}[t]
\vspace*{-2cm}
\resizebox{7.cm}{10.cm}{\includegraphics{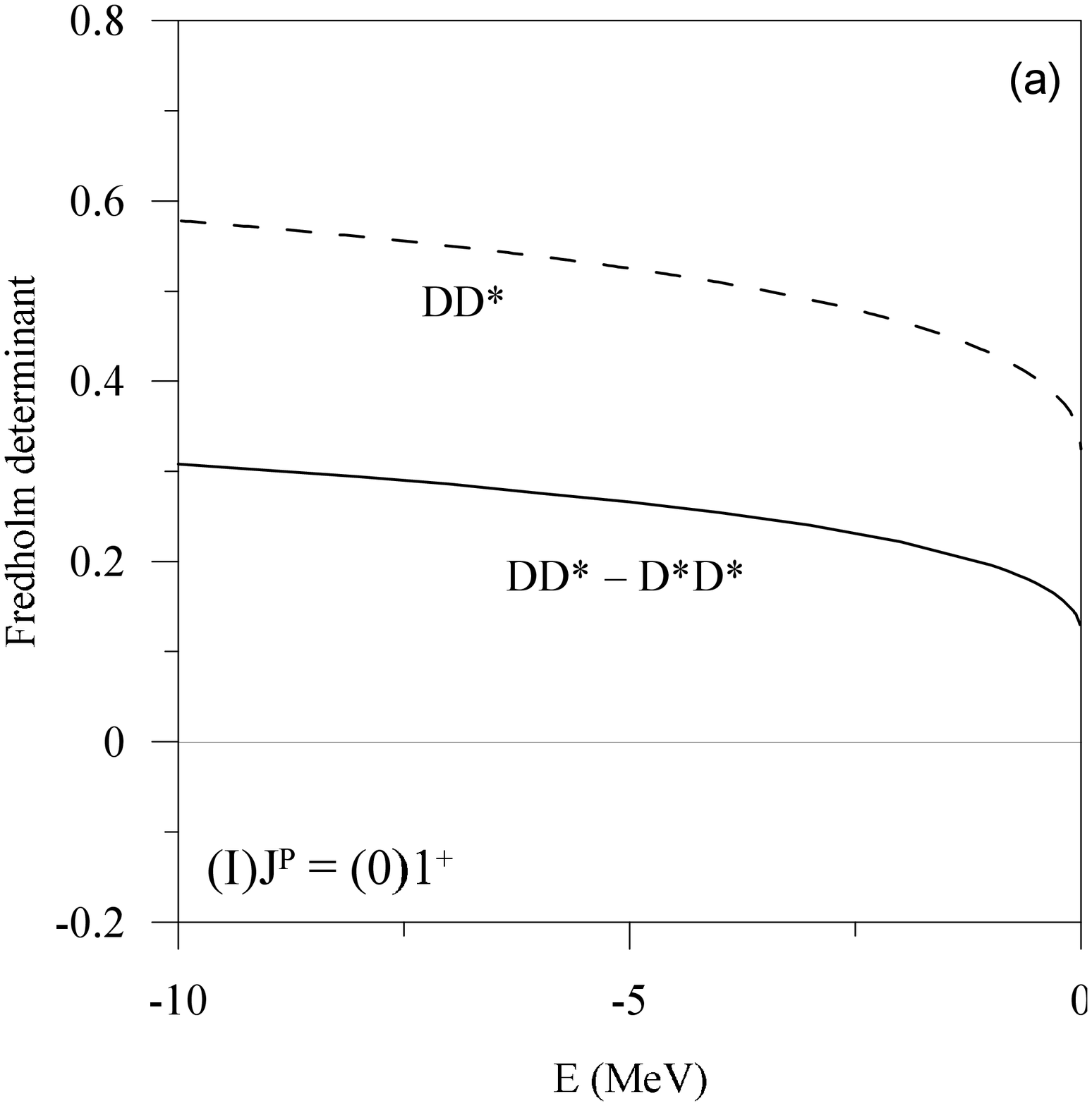}}\hspace{2cm}
\resizebox{7.cm}{10.cm}{\includegraphics{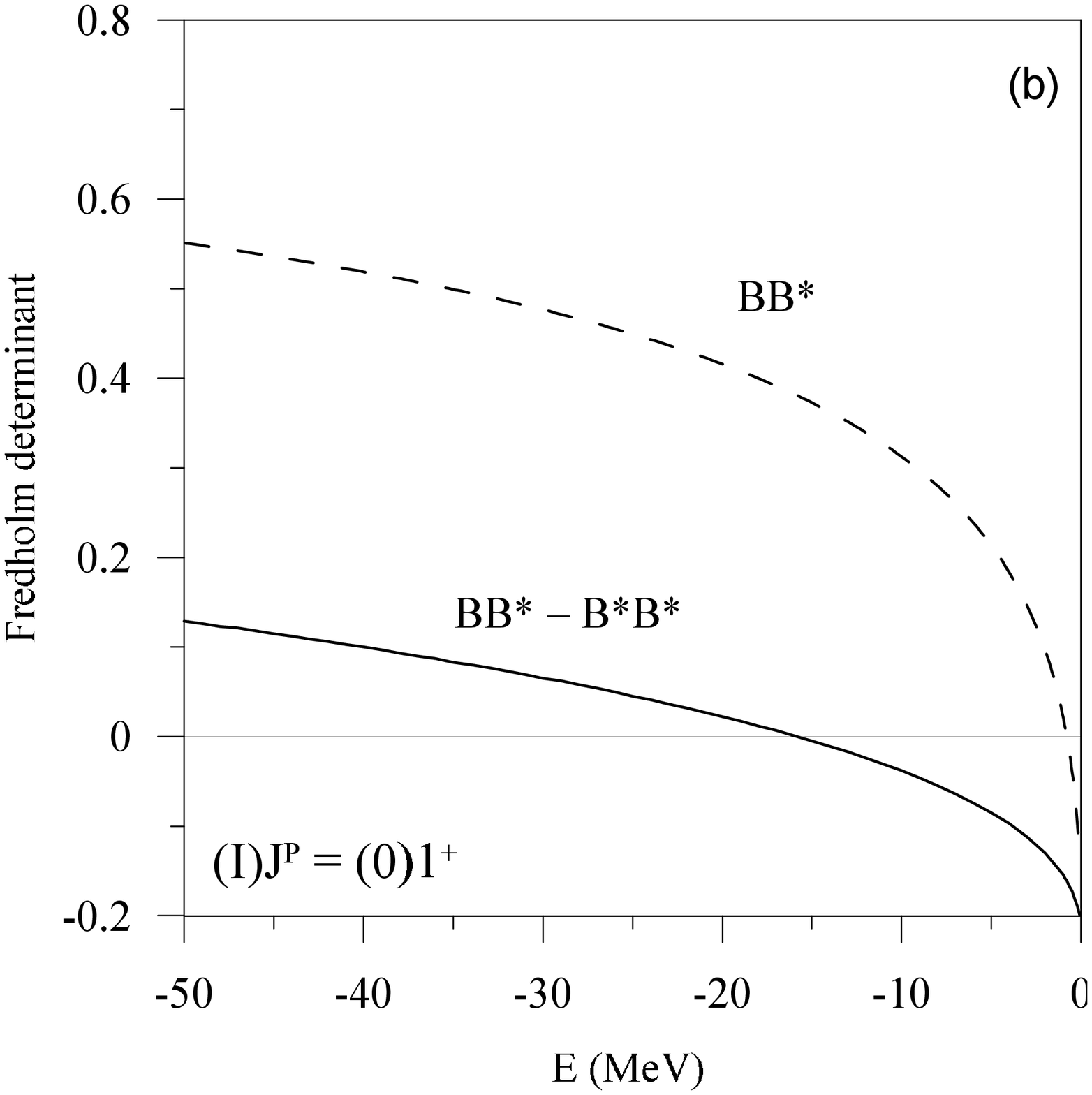}}
\vspace*{-2cm}
\caption{(a) $(I)J^{P}=(0)1^{+}$ $cc\bar n\bar n$ Fredholm determinant~\protect{\cite{Car11}}. The dashed line 
stands for a single channel calculation with the lightest two charmed mesons, the solid 
line includes the coupling to the relevant excited channels. (b) Same
as (a) for bottomonium.}
\label{fig3}
\end{figure*}

Thus, only a few channels may lodge molecular resonances. As discussed above, there is a remarkable exception to the general rule,
the $(I)J^{PC}=(0)1^{++}$ quantum numbers in the charmonium sector. In this case the 
$(c \bar n) - (n\bar c)$ ($D\bar D^*$) and $(c\bar c) - (n\bar n)$ ($J/\Psi \omega$) thresholds
are almost degenerate, and the attractive $D\bar D^*$ interaction together
with the cooperative effect of the almost degenerate two-meson thresholds
give rise to the widely discussed $X(3872)$~\cite{Car09}. 
In spite of the general idea that the stability of a system made of quarks
comes favored by increasing the mass of the heavy flavor, it becomes more complicated when several vectors in color
space contribute to generate a color singlet, as it it the case of four-quark systems~\cite{VijPR}. 
The reason is that, as explained above, the mass of one of the thresholds,
$(Q\bar Q) - (n\bar n)$, diminishes rapidly when the heavy quark mass increases (see Fig.~\ref{fig1}), making
therefore the meson-antimeson system, $(Q\bar n) - (n\bar Q)$, unstable. This simple reasoning
of coupled-channel calculations is illustrated in Fig.~\ref{fig2}.
The calculation is based in an interaction between quarks
containing a universal one-gluon exchange, confinement, and a chiral potential
between light quarks~\cite{Vij05}. As we can see in the left panel, the 
$D\bar D^*$ interaction (dashed line) is attractive, but not attractive 
enough to generate a resonance (the Fredholm determinant does 
not go through zero). It is the coupling to the almost degenerate $J/\Psi \omega$ 
channel (solid line) the responsible for having a bound state just
below the $D\bar D^*$ threshold. Such an explanation
comes reinforced by the recent observation of the decay $X(3872) \to J/\Psi \omega$~\cite{Amo10}.
When the mass of the heavy
quark is augmented from charm to bottom, the $B \bar B^*$ becomes more
attractive due to the decreasing of the kinetic energy and having
essentially the same interaction. However, the coupling to the lower channel,
$\Upsilon \omega$, destroys any possibility of having a bound state (see
right panel of Fig.~\ref{fig2}). Thus, based on the constituent quark
model ideas, one should not expect
a twin of the $X(3872)$ in bottomonium spectroscopy like those pointed out 
in hadronic models based on the traditional meson theory of the nuclear forces 
or resorting to heavy quark symmetry arguments~\cite{Tor92}.

One may also think that the proliferation of resonances above flavor thresholds 
could rely on our poor knowledge of confinement. Confinement is usually described in terms of simple pairwise 
interactions, but its realization at low energy may be much closer to a 
many-body force. Refs.~\cite{Car07} have analyzed the stability
of $Q n \bar Q \bar n$ and $Q Q\bar n \bar n$ systems in a simple string model by
considering only a multiquark confining interaction in an attempt
to discern whether confining interactions not factorizable as
two-body potentials would influence the stability of four-quark
states. The ground state of systems made of two quarks and
two antiquarks of equal masses was found to be below the
dissociation threshold. Whereas for the cryptoexotic $Q n\bar Q \bar n$
the binding decreases with increasing mass ratio $m_Q/m_n$,
for the flavor exotic $Q Q \bar n \bar n$
the effect of mass symmetry breaking is opposite. 
This effect contrary to binding will be even larger
when the mass ratio between the heavy and the light quark becomes larger,
as it would be the case of molecular structures in
the charmonium or bottomonium spectra.

The discussion on the last paragraph drives us to a brief comment on exotic states
$QQ \bar n \bar n$. In this case the situation is rather different to the nonexotic
$Qn \bar Q \bar n$ system, because the possible
dissociation thresholds do not contain
states made of a heavy quark and a heavy antiquark, whose binding would increase linearly
with the mass of the heavy flavor.
Thus, stability will be favored by increasing the mass of the heavy
flavor, being much more probable in the bottom sector than in the
strange one. This simple coupled-channel effect, that has been  discussed in detail in
Ref.~\cite{Car11}, is illustrated in Fig.~\ref{fig3}. As we can see,
the coupling to the heavier vector-vector two-meson state makes the system 
more attractive, the effect being much more important for the system containing 
two bottom quarks, generating thus larger binding energies. The search of such exotic
states is a hot experimental subject for the incoming years in different experimental
facilities~\cite{Exp11}.

Our discussion above may be easily generalized to the possible existence of charged
resonances contributing to charmonium or bottomonium spectroscopy. In this case, the
coupling to channels containing the light pion destroys the degeneracy between meson-antimeson
and charmonium-light two-meson thresholds, an important mechanism for binding four-quark states
in the $I=0$ sector. This excludes, for example, the existence of charged partners of the
$X(3872)$, as explained in Ref.~\cite{Car09}. Only one S-wave channel, the $J^{PC}=2^{++}$, where the
coupling to the charmonium-pion two-meson system is prohibited, may be candidate for lodging
a resonance close above the $D^* \bar D^*$ threshold, as has been already discussed both from the
experimental and theoretical points of view~\cite{Cha08}.

Summarizing, recent experimental data on charmonium spectroscopy have suggested the
existence of a large number of states above charmed meson thresholds.
They have been baptized as $X's$, $Y's$, and $Z's$, due to their unusual properties
not easily explained in terms of simple quark-antiquark pairs. Such proliferation of states
has pointed out to the existence of meson-antimeson molecules. In a quark-model picture we have justified
how such molecules may contribute to the light meson spectroscopy. In particular,
they could explain the existence of non quark-antiquark states for quantum numbers
that can be obtained from four-quarks in an S-wave but need orbital angular momentum from
a quark-antiquark pair. When increasing the mass of the heavy flavor, the possibility of
having meson-antimeson resonances decreases with the mass of the heavy quark. Only in some
particular cases the cooperative effect of nearby two-meson channels with an attractive
meson-antimeson interaction may produce resonances in the charmonium sector, the $X(3872)$ being the example
par excellence. Increasing the mass of the heavy quark destroys the possibility of a
twin state in the bottom sector, against the predictions of hadronic models based on the 
traditional meson theory of the nuclear forces or heavy quark
symmetry. Improved confinement interactions considering many-body forces do not
enhance the probability of having stable four-quark states
in the energy region close to the flavor thresholds. In fact, confining many-body forces
would go against the stability of non-exotic four-quark states. Finally, in the exotic sector, due 
to the nonexistence of thresholds made of two heavy quarks, the stability of two-meson states
would increase with the mass of the heavy quark.

The scenario that has been discussed in the present letter is far from taking 
your favorite random model, finding the $J^{PC}$ states that agree with it
and then ignoring, excusing or tweaking the model to apologize for those that 
do not. It is important to keep the big picture in mind if we are to progress.
Focus on the wood, not the trees. Only better experimental data will give the definite
answer to the correctness of the ideas drawn in this letter. The scenario proposed
would fit the current experimental data through the simple extension of the constituent
quark model we have presented and provides a smooth transition from quark to 
hadronic degrees of freedom. We hope that our contribution will stimulate a critical analysis
of the recent experimental data and theoretical investigations to disentangle resonances
from thresholds.

\acknowledgments
This work has been partially funded by the Spanish Ministerio de
Educaci\'on y Ciencia and EU FEDER under Contract No. FPA2010-21750,
and by the Consolider-Ingenio 2010 Program CPAN (CSD2007-00042).

\end{document}